\documentclass[prl,twocolumn,aps,showpacs]{revtex4}
\usepackage{amsfonts}
\usepackage{euscript}
\usepackage[english]{babel}
\begin{document}
\headsep 2cm

\title{
General relativity principle and uniqueness in Einstein equations}

\author{Mauricio Leston}\thanks{Electronic address: mauricio@iafe.uba.ar\\ }

\author{Rafael Ferraro}\thanks{Member
of Carrera del Investigador Cient\'{\i}fico (CONI\-CET,
Argentina).\\ Electronic address: ferraro@iafe.uba.ar\\ }

\affiliation{Instituto de Astronom\'{\i}a y F\'{\i}sica del
Espacio, Casilla de Correo 67, Sucursal 28, 1428 Buenos Aires,
Argentina,\\ and Departamento de F\'{\i}sica, Facultad de Ciencias
Exactas y Naturales, Universidad de Buenos Aires, Ciudad
Universitaria, Pabell\'on I, 1428 Buenos Aires, Argentina}


\begin{abstract}

The issue of implementing the principle of general relativity in
Einstein equations has been widely discussed, since Kretschmann's
well-known criticism stated that general covariance of the
Einstein equations is not suffice to express the principle of
general relativity (the equivalence of all the coordinate
systems). This failure is usually rooted in the fact that metric
in Einstein equations is not univocally determined by the matter
distribution. We show that the condition of univocal determination
of the metric by the matter distribution is stronger than the
requirement of equivalence of all coordinate systems. In order to
separate the uniqueness problem in Einstein equations from the
issue of the principle of general relativity, we define the
"equivalence group" instead of the notion of covariance group
which is empty of physical content. Moreover, we complement in a
positive way Kretschmann's objection by supplementing Einstein
equations with a sufficient condition for the equivalence of all
the coordinate systems. \vskip1cm

\end{abstract}


\maketitle

\section{1. Introduction}
Physics has a peculiarity that distinguishes it from any other
science: its theories are expressed in a language that differs
from the one in which they were developed. In the translation
process some generality may be lost or the original idea may end
up being contradicted. The latter is what happened to the
principle of general relativity in its translation to Einstein
equations. The aim of this work is to justify this assertion and
indicate the way to a more faithful translation.

As it is well known, Einstein felt attracted by Mach's thought
around 1912. Mach's criticism to Newton's absolute space, namely
his quoted comment regarding the rotating bucket, suggested to
Einstein and other authors the formulation of a new mechanics
where the inertial forces appearing in the so called non-inertial
frames would not be ascribed to the acceleration relative to the
absolute space but to the motion of the frame relative to the
matter distribution in the universe. From this point of view, the
force producing the parabolic shape of the surface of water in the
frame fixed to the bucket should be ascribed to the rotating
motion of the stars in that frame.

Initially, Einstein thought that Mach's ideas could be expressed
in a theory where the components of the metric and the affine
connection in each frame were obtained from the distribution of
matter of the universe. Certainly, the law linking the metric and
the distribution of matter should not privilege any frame.
Einstein convinced himself that this condition would be fulfilled
if the set of equations for the metric were generally covariant.

As Einstein recognized after Kretschmann's criticism
\cite{Kretschmann} (later clearly stated by Anderson
\cite{Anderson0, Anderson} and Friedman \cite{Friedman}), the
general covariance of the equations for the metric field does not
imply a principle of general relativity. In fact, the Newtonian
theory of gravitation can also be enunciated in covariant
language. The failure of the initial aspiration can be rooted in
the following fact: the metric in Einstein equations is not
univocally determined by the matter distribution. Einstein
equations constitute a system of differential equations for the
components of the metric tensor that, due to their differential
character, must be supplemented with additional conditions. Here
is the key to break the equivalence of all coordinate systems.
While the privilege of the inertial systems in the Newtonian
theory of gravitation formulated in covariant language comes from
imposing a particular form for the affine connection, the case of
Einstein equations is more complex.

Until now, no set of complementary conditions has been found which
may fix the components of the metric, and be compatible with the
principle of general relativity. As a consequence, we have a
theory in which the forces of inertia in the non-inertial systems
are attributed to a metric whose form is partly determined by
matter and partly  by conditions imposed from the  outside.
Consequently, Einstein's aspiration merely culminated in a metric
theory of gravitation, deserving the same criticism that Mach
addressed to Newtonian mechanics; absolute space was not
eliminated.

Because of the decisive role played by Mach as a trigger for
Einstein's ideas, many authors have considered both alternative
theories of gravitation and restrictions to the space of solutions
in Einstein equations, such that the initial aim can be reached
totally or partially. In the case of alternative theories that
retain a metric conception of the gravitation, the common goal is
the formulation of a theory where the metric is exclusively
determined by the matter distribution. The different meanings that
the expressions ``determination of the metric" and ``matter
distribution"  have for different authors gave rise to a wide
range of statements of the so called ``Mach principle". Bondi and
Samuel \cite{Bondi} mention ten different and non-equivalent
statements of the Mach principle. The Ref. (\cite{Barbour}) can be
consulted to have both a historical and an updated account of this
subject.

We will briefly comment the References \cite{Gilman, Sciama,
Lynden-Bell0}, which tackled the problem by proposing a relation
metric-distribution of matter which results from using a ``Green
function" for the metric tensor, analogously to electromagnetism
where the potential vector is expressed in terms of the current
source. With some differences, these works try to implement the
idea that the metric should be dictated by the distribution of
matter. Only some subgroup of metrics solving the Einstein
equations satisfies this requirement, which works as a selection
rule to eliminate ``non- Machian" solutions. Concerning the
implementation of the principle of general relativity, we should
point out that these works do not entirely reach their objective.
First, given the non-linear character of the Einstein equations,
the ``Green function" contains information of the same metric.
Therefore, the procedure is not a perfect translation of the
requirement of univocal determination of the metric by means of
the energy-momentum tensor. Second, the different formulations of
Mach's principle contain two intermingled questions: the
determination of the geometry by the matter distribution, and the
determination of the components of metric tensor by the form of
the matter distribution in a certain frame. As this paper will
show, the latter question, not the former, is the one involved in
the principle of general relativity. Our main claim in this paper
is that the condition of univocal  determination of the metric by
matter distribution is stronger than the requirement of
equivalence of all the coordinate systems dictated by principle of
general relativity.

Since our specific interest in this paper lies in the intersection
of Mach principle and the principle of general relativity, we will
focus exclusively on finding the requirements that a relation
metric-matter distribution should fulfill to satisfy the principle
of general relativity. In other words, we will try to find
selection rules that are a faithful translation of the principle
of general relativity.

The search for these rules has been guided by the following
principle. As usual, finding a solution to Einstein equations is
facilitated by the imposition of symmetries to the metric.
However, in other theories where a field is determined by its
source --for example, electrostatics-- the imposition of
symmetries of the source to the field solution is not merely a way
to solve the equations but a consequence of the very laws of the
field, as we will discuss in detail at the beginning of the next
section.

This observation leads to the idea of elevating to the category of
principle the association between symmetry of the source and
symmetry of the solution in Einstein equations. In order to put
this principle in practice, we consider that a supplementary
equation should be added to Einstein equations. This extra
condition on the solutions of General Relativity would cover part
of the path towards the generalization of the principle of
relativity.

In section 2 we analyze the electrostatic equations in order to
draw a parallel with Einstein equations. In section 3 we trace the
analogy between Einstein equations and the electrostatic ones. In
section 4 we analyze the problem of the ``determination of the
metric" by ``the matter distribution" by focusing on the meaning
of both expressions. In section 5 we examine the root of the
problem: the relation between the non-univocal determination of
the metric in Einstein equations by the energy-momentum tensor and
the general relativity principle. In section 6 we propose that a
supplementary equation be added to Einstein equations to express
the principle of general relativity. And we also show the results
of applying the supplementary equation to certain solutions
considered in the literature as ``non- Machian". Finally, we
present a summary and comment on the problems left open for a
future work.

\section{2. The electrostatic analogy}

\subsection{2.1 Covariance group and equivalence group}

To begin with, we will consider the electrostatic equations as a
toy model to study some aspects that should be also present in a
theory of gravitation. The electrostatic field $\vec{E}_C$
generated by a distribution of charge $\rho$ is

\begin{equation}
\vec{E}_C(\vec{r})=\int\frac{\rho(\vec{r'})
(\vec{r}-\vec{r'})}{{|\vec{r}-\vec{r'}|}^3}\
d^3r'\label{Ecoulombian}
\end{equation}
The Coulombian field $\vec{E}_C$  has  relational character
relative to the charge density; i.e., the differences between the
components of the field in two different Euclidean frames are
exclusively attributable to the different appearances of the
charge distribution in each frame. This means that the theory
given by ({Ecoulombian}) respects the isotropy and the homogeneity
of the space or, in other words, the equivalence of all Euclidean
frames. We will call GE, without distinction, both the set of
Euclidean frames or the group of transformations connecting them.

Let us consider a local consequence of Eq. (\ref{Ecoulombian}) by
taking divergence in both members:
\begin{equation}
\vec{\nabla}\cdot\vec{E}=\rho\label{E1}
\end{equation}
We call the theory given by this sole equation {\bf TE1}. Since
the mapping $\vec{E}\longrightarrow\vec{\nabla}\cdot\vec{E}$ is
non-injective, TE1 also contains other fields different from
(\ref{Ecoulombian}). So {\bf TE1} is a wider theory than the
previous one. Does {\bf TE1} respect the equivalence of the
elements of {\bf GE}? Although this seems to be a well posed
question, depending on what we understand by ``respects the
equivalence of the elements of {\bf GE}" we will get opposite
answers. If we are meaning that the fields $\vec{E}$ and $\rho$
satisfy the same equation in all the frames belonging to {\bf GE},
the answer is yes. This is a consequence of the conjunction of two
properties of the Eq. (\ref{E1}): its covariance under Euclidean
transformations and the absence of any reference to a particular
coordinate system \footnote[1]{Remarkably, the mere covariance of
the Eq. (\ref{E1}) does not imply the equivalence of the elements
of {\bf GE}. In fact the covariance group of Eq. (\ref{E1}) could
be increased by using the covariant derivative associated with a
flat connection $\Gamma$. In this case, the Eq. (\ref{E1}) would
acquire a covariant form under general coordinate changes.
However, the cost of increasing the covariance is the need to
introduce the connection $\Gamma$ as an additional object. In
other words, the equation (\ref{E1}) is identical in all the
coordinate systems  if written with $\vec{E}, \rho, \Gamma$; but
it is not if written with $\vec{E}$,$\rho$. Unless some additional
structure is introduced, the covariance group of Eq. (\ref{E1}) is
{\bf GE}.}.

As it was already said, the question admits another interpretation
which concludes that {\bf TE1} does not respect the equivalence of
the elements of {\bf GE}: the theory {\bf TE1} contains solutions
$\vec{E}$ which are related with the source $\rho$ in such a way
that some particular coordinate system turn out to be privileged.
In fact, the equation (\ref{E1}) admits both the solution
(\ref{Ecoulombian}) and the following non-relational solutions:
\begin{equation}
\vec{E_{\vec{a}}}(\vec{r})=\int\frac{\rho(\vec{r'})(\vec{r}-\vec{r'})}{{|\vec{r}-\vec{r'}|}^3}\
d^3r' + E_o\ \hat{a}\label{Ea}
\end{equation}

\begin{equation}
\vec{E_{\vec{j}}}(\vec{r})=\int\frac{\rho(\vec{r'})(\vec{r}-\vec{r'})}{{|\vec{r}-\vec{r'}|}^3}\
d^3r'+
\int\frac{\vec{j}(\vec{r'})\times(\vec{r}-\vec{r'})}{{|\vec{r}-\vec{r'}|}^3}\
d^3r'\label{Ej}
\end{equation}
where $E_o\ \hat{a}$ is an uniform arbitrary vector, and
$\vec{j(r)}$ is an arbitrary axial vector field. In contrast to
the relational field (\ref{Ecoulombian}), the fields (\ref{Ea})
and (\ref{Ej}) are non-relational regarding the source $\rho(\vec
r)$. In fact, the differences between components of the fields in
two different {\bf GE} frames cannot be exclusively attributed to
the appearance of the source $\rho$ in each Euclidean frame,
because they also depend on the appearance of $E_o\ \hat{a}$ and
$\vec{j}(\vec r)$ in each frame. In the laws (\ref{Ea}) and
(\ref{Ej})  the only equivalent systems are those related by
transformation that are symmetries of the vector $\hat{a}$ and
$\vec{j}$ respectively.

Summing up, the question of which coordinate systems are
equivalent in the theory given by (\ref{E1}) admits two different
readings. In the first case, we are asking about the group of
transformations that leaves the form of Eq. (\ref{E1}) unchanged.
This is the {\it covariance group} of Eq. (\ref{E1}). In the
second case, we are asking about the group of transformations
which does not  modify any injective mapping
$\rho\longrightarrow\vec{E}$ satisfying Eq. (\ref{E1}). This is
the covariance group common to all equations defining injective
mappings $\rho\longrightarrow\vec{E}$ that fulfills Eq.
(\ref{E1}).

The ambiguity in the issue of the equivalence of frames can be
eliminated by applying the notion of {\it equivalence group}
\footnote[2]{For a formal definition of related terms connected
with the covariance of a space-time theory, see Ref.
\cite{Friedman}.}. Let $\mathcal{D}_{\{y_i\}} (i=1.., N)$ be a set
of equations with variables $y_i$, and
$\mathcal{C}_{\mathcal{D}_{\{y_i\}}}$ their covariance group. We
defined the equivalence group $\mathcal{O}^{(y_1,y_2,..y_r|y_
{r+1}...,y_{n})}_{\mathcal{D}_{\{y_i\}}}$ of
${\mathcal{D}_{\{y_i\}}}$ as the covariance group shared by all
equations defining injective mappings $ (y_{r+1},...,y_{N})$ $
\longrightarrow (y_1,y_2,..y_r)$ that satisfies ${\mathcal
D}_{\{y_i\}}$. We call ``dynamical" the variables $y_1, ..., y_r$,
and ``initial conditions" the variables $y_{r+1}, ..., y_{N}$.

In this way, the issue of the equivalence group of a given theory
leads to different answers according to these two possible roles
of the variables entering an equation $\mathcal{D}$. We should
underline the difference between the concepts of equivalence group
and covariance group. As mentioned above, the latter is the set of
transformations preserving the form of a given equation; its
definition only depends on the form of the equation $\mathcal{D}$.
Its relation with the equivalence group can be written in the
following way:
\begin{equation}
\mathcal{C}_{\mathcal{D}_{\{y_i\}}}=\mathcal{O}_{\mathcal{D}_{\{y_i\}}}^{(..
|y_1,y_2,...,y_N)}\label{CO}
\end{equation}
where the right member displays all the variables in the role of
initial conditions.

Returning to example of theory TE1, three different equivalence
groups could be
considered:$\mathcal{O}^{(\vec{E}|\rho)}_{\mathcal{D}_{(\vec{E},\rho)}}$,$\mathcal{O}^{(\rho|\vec{E})}_{\mathcal{D}_{(\vec{E},\rho)}}$,
$\mathcal{O}^{ (..|\vec{E},\rho)}_{\mathcal{D}_{(\vec{E},\rho)}}$.
The different meanings of the question about the set of equivalent
frames in {\bf TE1} arise from the different roles the variables
$\rho$ and $\vec{E}$ can play. If we answer that all the frames of
{\bf GE} are equivalent, then we have asked about
$\mathcal{O}^{(..|\vec{E}, \rho)}_{\mathcal{D}_{(\vec{E},
\rho)}}$, where all the variables play the role of initial
conditions. In this case, the group of equivalence agrees with the
covariance group of the Eq. (\ref{E1}). If we answer that {\bf
TE1} does not respect the equivalence of the frames of {\bf GE},
then we have questioned on
$\mathcal{O}^{(\vec{E}|\rho)}_{\mathcal{D}_{(\vec{E}, \rho)}}$. In
this case $\vec{E}$ is a dynamical variable and $\rho$ is an
initial condition; then the equivalence group will be the
covariance group shared by all the injective mappings of $\rho$ in
$\vec{E}$ satisfying {\ref{E1}).
$\mathcal{O}^{(\vec{E}|\rho)}_{\mathcal{D}_{(\vec{E}, \rho)}}$ is
empty. In fact, the solutions (\ref{Ea}) and (\ref{Ej}) belong to
Eq. (\ref{E1}) and represent injective mappings of $\rho$ in
$\vec{E}$. The covariance group of each mapping are the rotations
around all axis parallel to $\hat{a}$ and the symmetries of
$\vec{j}$, respectively. Since $\hat{a}$ and $\vec{j}$ are
arbitrary, then the covariance group shared by all the mappings
(\ref{Ea}) and (\ref{Ej})is empty.

On the other hand, if the question refers to
$\mathcal{O}^{(\rho|\vec{E})}_ {\mathcal{D}_{(\vec{E}, \rho)}}$,
the answer is {\bf GE}, since the only injective mapping of
$\vec{E}$ (initial condition) in $\rho$ (dynamical variable)
satisfying the Eq. (\ref{E1}) is $\rho=\vec{\nabla}\cdot\vec{E}$.

The theory {\bf TE1} is a simple example of Kretschmann's well
known statement that -using the previously introduced language-
establishes that the covariance group of an equation is generally
smaller than the equivalence group of their solutions.
Nevertheless, there is a case where the covariance group of a
theory agrees with its equivalence group (besides the trivial case
(\ref{CO})). This happens when the equation $\mathcal{D}$ has a
unique solution for the chosen initial conditions (uniqueness
property).

\subsection{2.2 Equations for a relational field $\vec{E}$}

Although the equivalence group
$\mathcal{O}^{(\vec{E}|\rho)}_{\mathcal{D}_ {(\vec{E}, \rho)}}$
corresponding to Eq. (\ref{E1}) is empty, it is possible to
supplement the Eq. (\ref{E1}) with other equations, so that the
equivalence group of the extended system ${\mathcal D}'$ is equal
to some subgroup of {\bf GE}. For instance, we could get an
equivalence group $\mathcal{O}^
{(\vec{E}|\rho)}_{\mathcal{D}'_{(\vec{E},\rho)}}$ equal to the
covariance group $\mathcal{C}_{{\mathcal D}'}$ by adding equations
leading to the uniqueness of the solution for each initial
condition $rho$. In other words, certain conditions could be
imposed in order to restrict the domain of the functional
$\vec{E}\longrightarrow\vec{\nabla}\cdot\vec{E}$ so that it admits
the inverse mapping. Depending on the characteristics of these
conditions, the emerging theory will have different equivalence
groups which are subgroups of {\bf GE}.

For instance, the theory that will be called {\bf TE2}, which
results from adding to Eq. (\ref{E1}) the conditions
\begin{eqnarray}
\vec{\bigtriangledown}\times\vec{E}=0\label{E2a}\\
\lim_{\vec{r}\rightarrow\infty}\vec{E}(\vec{r}) = E_o\
\hat{a}\label{E3a}
\end{eqnarray}
have the field (\ref{Ea}) as the only solution associated to the
source $\rho$. It can be said that Eq. (\ref{E1}) has been
included in a theory whose equivalence group is a subgroup of {\bf
GE} containing the symmetries of the field $E_o\ \hat{a}$.

If we want to increase the equivalence group so that it agrees
with {\bf GE}, we can add to (\ref{E1}) the following conditions:
\begin{eqnarray}
\vec{\bigtriangledown}\times\vec{E}=0\label{E2}\\
\lim_{\vec{r}\rightarrow\infty}\vec{E}(\vec{r})=0\label{E3}
\end{eqnarray}
thus leading to the Coulombian Electrostatics for localized
sources. Because Eqs. (\ref{E1}), (\ref{E2},\ref{E3}) are
covariant under {\bf GE} and determine univocally the field
$\vec{E}$ for a given $\rho$, we can be sure that the equivalence
group of this system will agree with {\bf GE}. In fact, the
relational field (\ref{Ecoulombian}) is the only solution to Eqs.
(\ref{E1}), (\ref{E2},\ref{E3}). So, the conditions
(\ref{E2},\ref{E3}) eliminate the non-relational solutions
contained in Eq. (\ref{E1}).

The most general way to obtain that (\ref{E1}) be part of a theory
with an  equivalence group equal to {\bf GE} consists in adding an
equation that imposes the relational character to any field
solution. The relational character lies in the fact that the field
$\vec{E}({\vec r})$ must depend exclusively on the values of the
charge distribution $\rho$ and the position ${\vec r}$ relative to
each point where $\rho$ is non-null.

Let $\Lambda$ be a transformation belonging to the group of
rotations, translations and reflections. $\Lambda$ maps Cartesian
frames into Cartesian frames. Then, the relational character of
the field is expressed by the equation:
\begin{equation}\label{E4} \vec{E}^{(\rho^{\Lambda})}({\mathbb
M}_{(\Lambda)}\vec{r})={\mathbb
M}_{(\Lambda)}{\vec{E}}^{(\rho)}(\vec{r}) \end{equation}
${\vec{E}}^{(\rho)}$ being the field associated with the source
$\rho$, $\rho^{\lambda}$ the source transformed by $\Lambda$ and
${\mathbb M}_{(\Lambda)}$ the matrix representation of $\Lambda$
in the space of polar vectors. The theory given by the systems
(\ref{E1})-(\ref{E4}) will be called {\bf TER}.

In order to prepare the treatment of the Einstein equations, we
will rewrite the Eq. (\ref{E4}) in terms of the Lie derivative
$\mathcal{L}$. If $\Lambda$ belongs to the subgroup of continuous
transformations (translations and rotations) generated by a
vectorial field $\vec{\xi}$, -i.e. if the infinitesimal coordinate
transformation is ${\mathbb M}_{(\Lambda)}\vec{r} = \vec{r} +
\epsilon\ \vec{\xi} + {\cal O} (\epsilon^2)$-, then
$\rho^{\Lambda}\simeq \rho + \epsilon\mathcal{L}_{\xi}\rho$, and
the Eq. (\ref{E4}) is written as:
\begin{equation}
\vec{E}^{\rho +
\epsilon\mathcal{L}_{\xi}\rho}-\vec{E}^{\rho}\simeq\epsilon
\mathcal{L}_{\xi}\vec{E}^{\rho}\label{Elie}
\end{equation}
If the source admits some vector such that
$\mathcal{L}_{\xi}\rho=0$, then the Eq. (\ref{Elie}) adopts a
particularly simple form. This happens when the source has a
symmetry. By applying Eq. (\ref{Elie}) together with (\ref{E1}) to
the case of a spherically symmetrical charge $\rho$, there results
a Coulombian field (\ref{Ecoulombian}) fulfilling Eqs. (\ref{E2},
\ref{E3}).

Note that, given an arbitrary charge distribution, {\it the theory
{\bf TER} does not univocally determine $\vec{E}$ from $\rho$}. In
other words, the Eq. (\ref{E1}) posseses relational solutions
different to the field (\ref{Ecoulombian}). For instance, the
field:
\begin{equation}
\vec{E}_{non-linear}(\vec{r})= \vec{E}_C +
\frac{k}{V_D}\int_{\mathcal{D}}\vec{E}_C(\vec{r})d^3r,
\label{Enolineal}
\end{equation}
where $D=\{\vec{r}/\rho(\vec{r})\neq0\}$ and $k$ is an arbitrary
constant, is a relational solution of Eq. (\ref{E1}). In fact the
second term in Eq. (\ref{Enolineal}) is a constant field which
does not privilege any particular direction of the space, since
this term has been built exclusively from global properties of
$\rho$.

The theory {\bf TER} contains all the solutions of the extended
system that result from supplementing the Eq. (\ref{E1}) with the
equations:
\begin{eqnarray}
  \vec{\bigtriangledown}\times\vec{E}=\vec{\mathcal{J}}(\rho)\label{E5}\\
  \lim_{\vec{r}\rightarrow\infty}\vec{E}(\vec{r})=\vec{\mathcal{E}}(\rho)\label{E6}
\end{eqnarray}
for any vectors $\vec{\mathcal{E}}$ and $\vec{\mathcal{J}}$ built
exclusively from the charge distribution.

Nevertheless, the system (\ref{E1}) + (\ref{E4}), whose
equivalence group is {\bf GE}, acquires the uniqueness property by
additionally demanding that the field $\vec{E}^{\rho}$ be a linear
functional of its source, i.e., that $\vec{E}^{\rho_1 +
\rho_2}=\vec{E}^{\rho_1} + \vec{E}^{\rho_2}$ (this requirement is
feasible due to the linearity of Eq. (\ref{E1})). In this way, the
linear theory included in {\bf TErelacional} is equivalent to {\bf
TEC}. In fact, we already said that the Coulombian field is the
solution of Eqs. (\ref{E1}) + (\ref{E4}) for a spherically
symmetric charge distribution. This field satisfies the system
(\ref{E2}) + (\ref{E3}). The linearity implies that these same
conditions will be satisfied for all distribution $\rho$, since
all charge distribution can be expressed as a superposition of
spherically symmetric distributions. We have therefore proved that
the system (\ref{E4})-(\ref{E1}) supplemented with the linearity
requirement is equivalent to the system
(\ref{E1})-(\ref{E2})-(\ref{E3}).

To sum up, the equivalence group of {\bf TEC} is {\bf GE}. We have
defined the theory {\bf TE1} which is given by the sole equation
({\ref{E1}). The field (\ref{Ecoulombian}) belongs to {\bf TE1},
but this theory contains other solutions because {\it the map
$\vec{E}\longrightarrow\vec{\nabla} \cdot\vec{E}$ is
non-injective} and, therefore, it does not have inverse mapping.
The new theory has an equivalence group $\mathcal{O}^{
(\vec{E}|\rho)}_{\mathcal{D}_{(\vec{E},\rho)}}$ empty, although
its covariance is the same {\bf GE} as in the theory {\bf TEC}.
This reduction of the equivalence group happens because the space
of solutions of Eq. (\ref{E1}) contains both relational and
non-relational fields.

The Eq. (\ref{E1}) can appear in a theory whose equivalence group
is a subgroup of {\bf GE} (for instance the theory {\bf TE2} which
involves fields that privilege a particular direction of the
space). To render Eq. (\ref{E1}) a part of the theory {\bf TER}
-whose equivalence group includes all Euclidean transformations-
Eq. (\ref{E1}) was supplemented with the condition (\ref{E4}) that
imposes the relational character of the field.

Although the equivalence group of {\bf TER} is the set of all
Euclidean transformations, the field $\vec{E}$ is not univocally
determined  by the source $\rho$ in this theory. In other words,
the existence of the theory {\bf TER} proves that the solution of
Eq. (\ref{E1}) cannot be univocally fixed by demanding that the
equivalence group coincide with {\bf GE}. Besides we must demand
linearity, a requirement that has nothing to do with the issue of
the equivalence group here considered. In this way we have shown
that the requirement of univocal determination of the field by its
source is stronger than the requirement of equivalence of {\bf
GE}.

\section{3. The Einstein equations}
\subsection{3.1 The equivalence group and other definitions}

All that has been  said about Eq. (\ref{E1}) can be immediately
extended to Einstein equations:
\begin{equation}\label{Ein1}
 {\mathbb G}({\mathfrak g})=k{\mathbb T}
 \end{equation}
where $k$ is the Einstein constant, and ${\mathbb G}({\mathfrak
g})$ is the Einstein tensor associated with the Lorentzian metric
tensor ${\mathfrak g}$. In fact, one should replace ``Euclidean
transformations" with ``general coordinates transformations",
``charge distributions" with ``matter-energy distribution" and
``electric field" with ``metrics". Therefore, although the form of
Einstein equations is generally covariant, it is not guaranteed
that any relation metric-distribution of matter satisfying
Einstein equations will respect the equivalence of all coordinate
systems. General covariance of Einstein equations implies only
that the equivalence group $\mathcal{O}^{(..|{\mathfrak g},
\mathbb{T})}_{\mathcal{D}_{({\mathfrak g},\mathbb{T})}}$ (here
$\mathcal{D}$ are the Einstein equations) is the group of general
coordinate transformations, which will be called {\bf GG}.
However, this is not the desired property, as explained in the
introduction. We are interested in the equivalence group
$\mathcal{O}^{({\mathfrak
g}|\mathbb{T})}_{\mathcal{D}_{({\mathfrak g},\mathbb{T})}}$. Like
Eq. (\ref{E1}), Einstein equations could be part of an enlarged
system of equations with an arbitrary equivalence group. The
equivalence group depends on the supplementary conditions added
for univocally determining the metric. So, we must study the way
of reducing the domain of the functional
$\mathcal{G}:g_{ij}\longrightarrow G_{ij}({\mathfrak g})$ so that
the non-relational character be eliminated.

Although the electrostatic analogy can help us to tackle a more
involved problem, we should stress a fundamental difference
between Eq. (\ref{E1}) and Einstein equations. Eq. (\ref{E1}) was
derived from the relation field-source (\ref{Ecoulombian}), which
respects the equivalence of the Euclidean systems. Therefore, Eq.
(\ref{E1}) does contain the relational field, and only needs
proper conditions to select the relational field as its sole
solution. Moreover, the supplementary equations (\ref{E2},
\ref{E3}) were deduced from the knowledge of the relational field.
However, Einstein equations have not been deduced from any
relation metric-matter distribution respecting the equivalence of
all coordinate systems. Therefore, we do not have the guidance of
such a solution for finding supplementary conditions for Einstein
equations. Furthermore, we do not even know whether such solution
exists.

In spite of it, we have shown that the relational field
(\ref{Ecoulombian}) can be obtained by adding to Eq. (\ref{E1})
the condition (\ref{Elie}), which directly expresses the
relational character of the field ( i.e. the equivalence of the
Euclidean systems in the relation field-source) plus the
requirement of linearity. Omitting this last condition, the system
(\ref{E1})-(\ref{Elie}) has only relational solutions although it
lacks the uniqueness property. This will be our strategy to find
supplementary conditions for Einstein equations that express the
relational character of the metric.

Before moving on, we will explicitly separate the problem we have
posed from the Cauchy problem. The Cauchy problem is concerned
with the initial conditions which are needed to univocally
determine a solution of the equations. Once given the components
of the energy-momentum tensor on the manifold, we select a
solution of Einstein equations by prescribing the components of
the metric tensor and their normal derivatives (subject to the
constraint equations) on a certain space-like hypersurface.
Although this procedure does lead to a unique set of components of
the metric tensor, the solution obtained does not have, in
general, relational character. This is so because the Einstein
equations tell us the dynamics of the metric tensor, but say
nothing about the values of its components. This situation is
comparable to Newtonian dynamics, whose Galilean covariant laws
cannot regulate the non-invariant velocities of the bodies.
However, we intend to get a theory where the metric, not just its
dynamics, is dictated by the distribution of matter. This plan
retains Einstein's original ideas motivated by Mach.

\subsection{3.2 The distribution of matter}

The matter distribution is characterized by the energy-momentum
tensor ${\mathbb T}$ in the right member of Eq. (\ref{Ein1}). We
are not interested here in the structure of the energy-momentum
tensor but only in its value at each point of space-time. Since
$\mathbb{T}$ is in the image of the functional $\mathcal{G}$
mapping $g_{ij}$ in $G_{ij}$, then the divergence of the
energy-momentum tensor is identically null. This divergence
involves the non-predetermined metric affine connection associated
with the pre-image of $\mathbb{T}$.

\subsection{3.3 The distinguishability of coordinate systems}

In the context of Electrostatics, different Euclidean frames can
be distinguished by the appearance of the charge in each one of
them. If a relational electrostatic field looks different in two
different Euclidean frames, then the charge looks different in
each frame. Analogously, a non-null energy-momentum tensor
$\mathbb{T}$ provides a physical way of distinguishing different
coordinate systems. Although different coordinate systems can be
distinguished by means of the distribution of energy-matter, this
does not mean that they are non-equivalent.

\subsection{3.4 The metric tensor}\label{mt}

The points --the events-- in space-time can be identified by means
of different coordinate systems. Thus, a given geometry
$\mathfrak{g}$ can be represented by different values of its
components at each point $\bf x$: $g_{ij}({\bf x})\equiv
<\mathfrak{g},{\bf e}_i\otimes{\bf e}_j>({\bf x})$, where $\{{\bf
e}_i({\bf x})$ is a coordinate basis at ${\bf x}$. For instance,
we could introduce a flat geometry by giving any of the equivalent
matrices $g_{ij}({\bf x})$ corresponding to a flat geometry. In
this case, each choice for $g_{ij}({\bf x})$ amounts to the choice
of the transformation connecting the basis $\{{\bf e}_i({\bf x})$
with the Euclidean basis. From a physical point of view, these
different ways of introducing a geometry in space-time are
undistinguishable, unless the different coordinate systems are
physically distinguishable. The energy-momentum tensor
$\mathbb{T}$ allows to distinguish coordinate systems, as it was
already stated.

In principle, any space-time could be provided with a metric not
subject to fulfilling the Einstein equations. Even so, the
procedure to introduce the metric involves two steps. First, a
geometry is chosen and written as one of the equivalent matrices
$g_{ij}$, and second, the basis $\{{\bf e}_i({\bf x})$ where the
metric has the components $g_{ij}$ is physically identified by
giving the components of $\mathbb{T}$ in that basis or the
transformation linking $\{{\bf e}_i({\bf x})$ with some privileged
basis selected by $\mathbb{T}$.

\section{4. The determination of metric in Einstein equations}

The Einstein equations do not determine the metric $\mathfrak{g}$
exclusively from the energy-momentum tensor $\mathbb{T}$. In fact,
the functional $\mathcal{G}$ mapping $g_{ij}$ in $G_{ij}$ {\it is
non-injective}, i.e., there exists $g_{ij}\neq {g'_{ij}}$ so that
$\mathcal{G}(g_{ij})=\mathcal{G} (g'_{ij})$. Therefore, the
functional $\mathcal{G}$ lacks of inverse mapping, unless its
domain is properly restricted. However, our goal is to study a
more weaker condition: the reduction on the domain in order to
eliminate the non-relational metric.

In order to study the reduction of the domain, the above mentioned
aspects of the specification of the metric should be reexamined
due to the relation between metric and energy-momentum tensor
established by the Einstein equations. Three sets are involved in
this issue:

\begin{tabular}{llc}
$\mathcal{A}_{(g_{ij})}=\{matrices \
g_{ij}/\mathcal{G}(g_{ij})=G_{ij}\}$
\\
$\mathcal{B}_{(G_{ij})}=\{geometries\ {\mathfrak{g}}\ included\
in\ \mathcal{A}_{(G_{ij})}\}$
\\ $\mathcal{C}_{(\mathfrak{g},
G_{ij})}=\{matrices\ g_{ij}\in \mathcal{A}_{(G_{ij})}\ associated\
with$
\\
$a\ given\ geometry\ {\mathfrak{g}}\in \mathcal{B}_{(G_{ij})}
\}\subset\mathcal{A}_{(G_{ij})}$
\end{tabular}
We notice that the set $\mathcal{C}_{(\mathfrak{g}, G_{ij})}$ does
not contain all the equivalent forms of geometry $\mathfrak{g}$
but only those satisfying the Einstein equations for a certain
expression $G_{ij}$ of the tensor $\mathbb{G}$.

Our aim is to replace the set $\mathcal{A}_{(G_{ij})}$ by a subset
$\mathcal {A}'_{(G_{ij})} \subset \mathcal{A}_{(G_{ij})}$
containing only one representation of each geometry. In other
words, $\#\mathcal{C}'_{(\mathfrak{g}, G_{ij})}=1$. This process
can be accomplished in two steps: i) an element is chosen from
$\mathcal{B}_{(G_{ij})}$; ii) an element is chosen from
$\mathcal{C}_{(\mathfrak{g}, G_{ij})}$. The first step only
concerns geometric aspects; then, it respects the equivalence of
all coordinate systems.  For instance, we consider the Minkowski
metric and the exact plane wave \cite{BPR}. They are two different
geometries associated with the null energy-momentum tensor. A
choice of some of them in the set $\mathcal{B}_{(G_{ij}=0)}$ does
not privilege any coordinate system.

The choice in step (ii) is the one that produces a conflict with
the principle of general relativity. Since it is involved with the
coordinate system, that choice must be dictated by the only
quantity associated with a given coordinate system: the components
$G_{ij}$ of $\mathbb{G}$. A clear example of this conflict is the
case where $G_{ij}$ is null in a finite region of space-time. In
fact, since general coordinate changes are local, we can build two
coordinate bases $\{\bf e(x)\}$ and $\{\bf e'(x)\}$ such that they
agree at all point of space-time, except in the empty region
$\mathcal{V}$ . Therefore, the components of the energy-momentum
tensor agree in the whole space-time:
\begin{equation}
<\mathbb{T},{\bf e_i}\otimes{\bf e_j}>({\bf x})=<\mathbb{T},{\bf
e'_i}\otimes{\bf e'_j}>({\bf x})\ ,\ \ \ \forall{\bf x}
\end{equation}
In spite of this, the components of the metric tensor will differ
in the region where the coordinate bases do not agree:
\begin{equation}
<\mathfrak{g},{\bf e_i}\otimes{\bf e_j}>({\bf
x})\neq<\mathfrak{g},{\bf e'_i}\otimes{\bf e'_j}>({\bf x})\  ,\ \
\ \ {\bf x}\in {\mathcal{V}}
\end{equation}
The sole acceptable changes of a relational metric tensor, when
evaluated in different coordinate systems, should be those arising
from changes of the components of the energy-momentum tensor in
some region of space-time.

\section{5. Obstacles to implementing a principle of general relativity}

In the former section we showed the existence of non-relational
metrics in Einstein-equation: those whose Einstein tensor
$\mathbb{G}(\mathfrak{g})$ is null in a finite region of
space-time. The non-relational character of this kind of metric is
clear: the values of the components $g_{ij}=<\mathfrak {g},{\bf
e_i}\otimes{\bf e_j}>({\bf x})$ of a metric $\mathfrak{g}$ in two
coordinate systems differing only inside the region where
$\mathbb{T}$ is null, cannot be attributed to differences in the
appearances of matter distribution because the components of
$\mathbb{T}$ are identical in both systems. As a corollary, a
distribution of matter compatible with the principle of general
relativity must ``fill" all the space-time. Evidently, the
requirement of an everywhere non-null energy-momentum tensor is
produced by the interest in having a theory with a general
equivalence group which includes local transformations. This
requirement would not be needed in a theory with a global
equivalence group (in the sense that its transformations would not
depend on ${\bf x}$, as it happens with the Cartesian
transformations in Electrostatics). In fact, in that case it would
be impossible to perform a coordinate change in the empty region
$\mathcal{V}$ without simultaneously modifying the components of
the energy-momentum tensor in the region where they are non-null.

Actually, the Einstein equations are far from being a pure
expression of the principle of general relativity in two ways. On
the one hand, the mere covariance of their form is not enough to
express the principle. On the other hand, the Einstein equations
govern the geometry, which is a property that has nothing to do
with coordinates. Therefore, both steps mentioned in Section
\ref{mt} are entangled in Einstein equations. This fact hinders
the analysis of the problem we are interested in, which is the one
corresponding to step (ii). A theory straightforwardly expressing
the equivalence of all coordinate systems should only establish
that the choice of $g_{ij}$ in step (ii) depend exclusively on the
matter distribution given by the components of the energy-momentum
tensor in the considered coordinate system. This would express
Mach's idea that the inertial forces in each frame are exclusively
determined by the motion of the frame relative to the matter of
the universe. But Einstein not only wanted to formulate the
principle of general relativity but also geometrize the
gravitational field, which led to the entanglement of the issues
considered in steps (i-ii).

\section{6. A supplementary condition in a theory with an general
equivalence group}

In the light of the analysis of Electrostatics made in Section 2,
it is natural to consider supplementing the Einstein equations
with a condition analogous to Eq. (\ref{Elie}). The supplementary
condition should avoid those solutions privileging coordinate
systems beyond the natural privilege that can be induced by the
source. Then, the supplementary condition should reduce the domain
of the functional $\mathcal{G}$ by imposing a dependency between
the components of $\mathfrak{g}$ and the components of
$\mathbb{T}$. As in Electrostatics, this procedure would not lead
to a unique relation metric-source. Let us denote with $
g_{ij}^{(T_{ij})}$ the components of a metric fulfilling the
supplementary condition. We will call this metric relational.
Following the electrostatic example, the relational character will
be expressed by an equation similar to Eq. (\ref{Elie}). Since we
are going to implement a general equivalence group, we must
replace the vectors $\xi$ generating Euclidean transformations by
others generating arbitrary coordinate changes. Therefore we
propose supplementing the Einstein equations with the condition
\begin{equation}\label{propuesta}
g_{ij}^{(T_{rs}^{(\varepsilon\xi)})}-g_{ij}^{(T_{rs})}\cong
\varepsilon({\mathcal L}_{\xi}g^{(T_{rs})})_{ij}
\end{equation}
where $\cong$ means that the equality is true in first order in
$\varepsilon$, and $T_{rs}^{(\varepsilon\xi)}\equiv{T}_{rs}+
\epsilon(\mathcal{L}_{\xi} \mathbb{T})_{rs}$. While
$\varepsilon({\mathcal L}_{\xi}g^{(T_{rs})})_{ij}$ is the change
of components of the metric under the infinitesimal coordinate
change generated by the vector $\xi$, the left member of Eq.
(\ref{propuesta}) compares two solutions of Einstein equations
associated, respectively, with the sources
$T_{rs}^{(\varepsilon\xi)}$ and $T_{rs}$. However,
$T_{rs}^{(\varepsilon\xi)}$ are nothing but the components of
$\mathbb{T}$ in the transformed coordinate system.

We cannot state that the Eqs. (\ref{Ein1}-(\ref{propuesta})
univocally determine the metric from the distribution of matter.
But this point has nothing to do with the principle of general
relativity. The Eqs. (\ref{Ein1})-(\ref{propuesta}) could contain
different relations metric-source, even those corresponding to
different geometries. This feature is also present in the Eqs.
(\ref{E1})-(\ref{Elie}), which admit solutions such as
(\ref{Enolineal}). A different solution to (\ref{E1})-(\ref{Elie})
for a same source $\rho$ differs in properties which are invariant
under Euclidean transformations. Instead, we claim that, once the
geometry has been fixed, Eq. (\ref{propuesta}) carries out the
additional desired work: it turns injective the relation {\it
components of $\mathbb{T}$ - components of $\mathfrak{g}$}. In
fact, Eq. (\ref{propuesta}) establishes that different components
of the metric are only ascribable to different components of the
energy-momentum tensor $\mathbb{T}$. Thus, Eq. (\ref{propuesta})
reduces the domain of $\mathcal{G}$ to accomplish the principle of
general relativity. We compare Electrostatics and Gravitation in
Table 1.

Although the Eq. (\ref{propuesta}) is equivalent to Eq.
(\ref{Elie}), we cannot directly pass to the analogous of Eq.
(\ref{E2})-(\ref{E3}) because the Einstein equations are not
linear. This fact hinders the detailed analysis of the content of
Eq. (\ref{propuesta}). However, two corollaries of Eq.
(\ref{propuesta}) are evident:

1) The Eq. (\ref{propuesta}) excludes those distributions of
matter that are null in some space-time regions. In fact, if
$\mathbb{T}$ was null in a region of space-time, we could
arbitrarily choose vectors $\xi$ being different from zero only in
that region. In this case, $\mathbb{T}^{ (\varepsilon\xi)}$ would
be equal to $\mathbb{T}$ in the whole space-time. Thus, Eq.
(\ref{propuesta}) would imply a null metric.

2) If the vector field $\zeta$ is associated with a symmetry of
${\mathbb T}$, i.e., $\mathcal{L}_{\zeta}(\mathbb{T})=0$, then
$\mathcal{L}_{\zeta} (\mathfrak{g})=0$. In this way the Eq.
(\ref{propuesta}) raises to the rank of principle the conditions
implicitly contained in the catalogue of symmetric solutions for
Einstein equations. Moreover, Eq. (\ref{propuesta}) regulates not
only symmetric solutions but even those solutions without
symmetries. We will now examine the role played by Eq.
({\ref{propuesta}) in some known solutions of General Relativity.

\subsection{6.1 The  exterior and interior Schwarzschild metric}
The exterior Schwarzschild metric can be regarded as the external
solution for a spherically symmetric matter distribution subject
to the asymptotically flat condition. For this reason it is
mentioned as an example of non-Machian metric, since it is
determined partly by the source and partly by the boundary
condition at the infinite. In fact, according to some definitions
of Mach principle, the geometrical properties of the metric must
be dictated exclusively by the matter distribution.

Nevertheless, concerning the principle of general relativity, this
aspect of the selection of the metric does not cause any trouble
because the asymptotically flat boundary condition is a
geometrical property (it does not privilege any particular
coordinate system).

The true conflict between the exterior Schwarzschild metric and
the principle of general relativity is in the fact that this
metric corresponds to an energy-momentum tensor which is null in
the outer finite region $r>R$ (where $R$ is Schwarzschild's
radius, and $r$ is the standard radial coordinate). Due to this
feature, this solution is ruled out by (\ref{propuesta}). In other
words, the coordinate system where the metric asymptotically
adopts the standard Minkowskian form should be determined by the
appearance of the matter distribution in it. However, such desired
state cannot be reached because the energy-momentum tensor is null
in the exterior region.

Instead, the symmetry of the interior metric is not implied by
Einstein equations but imposed by Eq. (\ref{propuesta}). If the
spherically symmetric source fills space-time, then the interior
solution satisfies the two necessary conditions involved in Eq.
(\ref{propuesta}). The form of this solution is not completely
fixed by the source; we can apply the Birkhoff theorem here. But
this non-uniqueness aspect of the metric does not cause any a
priori conflict with the principle of general relativity.

\subsection{6.2 The Reissner-Nordstrom metric}

The Reissner-Nordstrom metric is the outer solution for a
spherically symmetric charged object. As such, it is a good
example of a metric that satisfies both necessary conditions
involved in Eq.(\ref{propuesta}: the energy-momentum is non-null
in all space-time, and the metric shares the symmetry of the
source. Unlike the exterior Schwarzschild metric, the
asymptotically flat boundary condition does not exhibit, a priori,
a conflict with the principle of  general relativity. The
difference resides in the fact that there is a non-null matter
distribution which distinguishes the coordinate systems in the
asymptotic region.

\subsection{6.3 The metric associated to a rotating mass shell}
Brill and Cohen have studied the approximate solution to Einstein
equations corresponding to a rotating mass shell \cite{Brill1,
Brill2, Brill3}, and Pfister and Braun extended the approximation
to a higher order \cite{Pfister1, Pfister2}. Actually, Eq.
(\ref{propuesta}) rules out this kind of solutions, because the
space-time is empty outside the shell. Nevertheless, we could
consider a different solution in which a distribution of matter is
added in the empty region, being sufficiently weak and
asymptotically null, so that the metric is slightly changed. We
could even consider the solution found in Ref.
[\cite{Lynden-Bell}], where rotating concentric shells with
variable densities fill space-time. For our purpose, it will be
enough to think in the simpler approximate Brill \& Cohen
solution, which consists in a perturbation of the (non rotating)
Schwarzschild solution. They consider a rotating dust shell
localized at $r = R$, whose four-velocity field is:
$u^{\theta}=0=u^{r}$, $u^{\varphi} = u^0\, \omega(t)$, where $u^0$
does not depend on $\theta$ and $\varphi$. Brill \& Cohen propose
the outer solution
\begin{eqnarray}
ds^2&=-{(\frac{2r-m}{2r+m})}^2d{t}^2 + (1+\frac{m}{2r})^4[dr^2
\nonumber\\ &+ r^2 (d\theta^2 + sen^2\theta(d\varphi-\Omega(r,
t)dt)^2)]\label{Brill}
\end{eqnarray}
which can be regarded as a perturbation of Schwarzschild's
solution in isotropic coordinates. This metric is subject to the
following requirements:

1. It must fulfill Einstein equations in first order in $\Omega$
and $\omega$.

2. It can be matched with the flat interior metric, regarding the
junction conditions associated with the presence of the shell at
$r=R$.

3.The metric must be asymptotically Minkowskian. Then,
$\lim_{r\rightarrow\infty}\Omega(r, t) = 0$.

The inner metric is
\begin{eqnarray}
ds^2=&-{(\frac{2R-m}{2R+m})}^2d{t}^2 + (1+\frac{m}{2R})^4[dr^2 +
r^2(d\theta^2 \nonumber\\  &+
sen^2\theta(d\varphi-\Omega_o(t)dt)^2)]\label{Brillin}
\end{eqnarray}
where $\Omega_o (t)$ is determined by junction conditions. In Eq.
(\ref{Brillin}), the form $d\varphi-\Omega_o(t)dt$ is exact
($d\varphi-\Omega_o(t)dt$ $= d\varphi-\int \Omega_o(t) dt$ $=
d\varphi^{\prime}$). So the inner metric can adopt the Minkowskian
form by means of a coordinate change. It is evident that the
distribution of matter admits a coordinate system where it looks
spherically symmetric. However, the metric in this coordinate
system could not reflect the symmetry of the source because
$d\varphi-\Omega(t,r)\\dt$ is not an exact 1-form. In other words,
the only outer spherically symmetric solution of Einstein
equations is Schwarzschild's solution, but the metric
(\ref{Brill}) cannot be transformed to the Schwarzschild form.
Therefore, Brill \& Cohen's solution is not compatible with the
supplementary Eq. (\ref{propuesta}).

We should add that the former problem becomes more complex when
contributions of higher order in $\Omega$ are considered. In that
case, the rotation of the shell cannot be rigid but the speed
$\omega$ must depend on $\theta$. \cite{Pfister1, Pfister2}
Although the spherical symmetry of the source cannot be retained
for higher orders in $\Omega$, the supplementary Eq.
(\ref{propuesta}) would be enough to rule out this kind of
solution in the lowest order. Since the supplementary requirement
(\ref{propuesta}) could be applied to each order, then we can
state that this kind of solution is not relational in the sense
above defined.

\section{7. Summary}

We have exploited the electrostatic simile to study the way of
implementing a principle of general relativity. This strategy
begins by analyzing the additional conditions that we should add
to the Eq. (\ref{E1}) in order to rise its covariance group -the
set of Euclidean transformations- to the rank of equivalence
group. Since (\ref{Elie}) demands that solutions to Eq. (\ref{E1})
must be relational with its source, the equivalence group of the
extended system (\ref{E1})- (\ref{Elie}) is the Euclidean
transformations group. Nevertheless, if we do not impose the
condition of field linearity of the field with respect to its
source, the system (\ref{E1}-\ref{Elie}) does not contain a unique
relation field-source. Thus, this discussion allowed us to
distinguish between the requirement of univocal determination by
the source and the requirement of equivalence of all Euclidean
systems in the relation field-source. After imposing the
requirement of linearity with respect to the source, these two
requirements agree. In such case, the system (\ref{E1})
+(\ref{Elie}) turns out to be equivalent to the system
(\ref{E1}-\ref{E2}-\ref{E3}), whose sole solution is the
Coulombian field (\ref{Ecoulombian}).

In Einstein equations, the univocal determination of the metric by
the matter distribution is a condition stronger than the
requirement of equivalence of all coordinate systems dictated by
the principle of general relativity. We have shown in section 3.4
that there are two steps in the specification of a given metric.
The principle of general relativity implies that the choice in the
step (ii) should be dictated by the matter distribution, the
choice in the step  (i) being arbitrary. For this purpose, we have
supplemented Einstein equations with the condition
(\ref{propuesta}), which  says that the metric solution to Eq.
(\ref{E1}) should be relational. In this way, the equivalence
group of the extended system (\ref{Ein1}-\ref{propuesta})is the
set of all coordinate transformations.

This so extended system of equations contains two straightforward
general results:

I) The allowed energy-momentum tensor cannot be null in any region
of the space-time. Therefore Mach's criticism to the Newton bucket
experiment should be reformulated by demanding that the parabolic
form of the water surface is due to the motion relative to the
electromagnetic radiation filling the universe.

II) The symmetries of the matter distribution are also symmetries
of the metric. Although it is well known that symmetric solutions
of Einstein equations do have this property, this is not a
mandatory consequence of Einstein equations. The Eq.
(\ref{propuesta}) shows that this association is a consequence of
applying the principle of general relativity.

Then, we have given a positive complement to Kretschmann's
objection: the  Eq. (\ref{propuesta}) replaces the requirement of
general covariance which is empty of physical content. In spite of
the effort required for a more exhaustive analysis of the content
of Eq. (\ref{propuesta}), we can state that Eq. (\ref{propuesta}),
together with Einstein equations, implements the principle of
general relativity in the so called Theory of General Relativity.

\widetext

\begin{table}[h]

\vskip 1cm

\begin{tabular}{|l|c|c|} \hline
& &\\ & {\bf Electrostatics}&{\bf Gravitation}\\ &  & \\ \hline
 &  & \\
 \hskip.4cm {\bf I}.- Differential equation    & $\vec{\nabla}\cdot\vec{E}=\rho$ &
               ${\mathbb G}({\mathfrak g})=k{\mathbb T}$ \\
  &  & \\
 \hline
  &  & \\
 \hskip.4cm Covariance group   &\parbox{2in}{
   Euclidean transformations
    }&\parbox{2in}{general coordinate changes}\\
     &  & \\
 \hline
  &  & \\
\hskip.4cm Examples of non-relational solutions &
\parbox{2in}{
$\vec{E_{\vec{a}}}(\vec{r})=\int\frac{\rho(\vec{r'})(\vec{r}-\vec
{r'})}{|\vec{r}-\vec{r'}|^3}\ d^3r' + \vec{E}_o\ {\hat a}$ }
&\parbox{2in}{$\ \ \ \ \ \ \ \ $any solution containing$\ \ \ \ \
\ \ \ \ $ an empty region}
\\
 &  & \\
\hline
 &  & \\
\hskip.4cm {\bf II}.- Supplementary equation imposing & &\\
\hskip.4cm an equivalence group equal to the & $\vec{E}^{\rho +
\epsilon\mathcal{L}_{\xi}\rho}-\vec{E}^{\rho} \simeq\epsilon\
\mathcal{L}_{\xi}\vec{E}^{\rho}$&\parbox{2in}{
$g_{ij}^{T_{rs}^{(\varepsilon\xi)}}-g_{ij}^{T_{rs}}\cong
\varepsilon({\mathcal L}_{\xi}g^{T_{rs}})_{ij}$}\\ \hskip.4cm
covariance group of the equation (I) & &\\
 &  & \\
\hline
 &  & \\
\hskip.4cm Examples of injective mappings &\parbox{2in}{
$\vec{E}_C(\vec{r})=\int\frac{\rho(\vec{r'})(\vec{r}-\vec{r'})}{|\vec{r}-\vec{r}'|^3}
\ d^3r'$ }
 & not found yet\\
 \hskip.4cm fulfilling (I) + (II)& &\\
  &  & \\
 \hline
 &  & \\
\hskip.4cm {Non-uniqueness of the solutions of (I-II)$\ \ \ $ }
 &\parbox{2in}{(II) is equivalent to:
\begin{eqnarray}
\vec{\bigtriangledown}\times\vec{E}=\vec{\mathcal{J}}(\rho)\nonumber\\
\lim_{\vec{r}\rightarrow\infty}\vec{E}(\vec{r})=\vec{\mathcal{E}}(\rho)\nonumber
\end{eqnarray}
for all vectors $\vec{\mathcal E}$ and $\vec{\mathcal J}$ built
exclusively from the charge distribution. There are many vectors
$\vec{\mathcal E}$ and $\vec{\mathcal J}$ for a given distribution
$\rho$} &\parbox{2in}{existence of different geometries associated
to the same tensor $\mathbb G$ }\\
 &  & \\
\hline
\end{tabular}
\centering \caption{ \it Comparison between Electrostatics and
Gravitation}
\label{extab}
\end{table}

\narrowtext

\end{document}